\documentclass[12pt]{article}
\usepackage{openwork}
\usepackage[dvipsnames]{xcolor}

\title{Towards Supporting Quantum Grassroots Organizations in ASEAN}
\author[1]{Amalina Lai}
\author[1]{Bethel Loke Hui Ting}
\affil[1]{Singapore Quantum Youth Collective, Singapore}
\date{}

\begin{document}

\maketitle

\begin{abstract}
This paper investigates the role of quantum grassroots organizations in The Association of Southeast Asian Nations (ASEAN), the challenges they face, and how these organizations can be better supported by other players in the ecosystem of quantum technologies. We look into the goals and strategies of ASEAN grassroots organizations and their potential impacts and investigate the obstacles and challenges currently faced by these groups. We propose providing mentors with specialized knowledge to assist in the refinement of technical content and organizational management, building connections between grassroots and key stakeholders in the quantum space, and creating opportunities to obtain stable funding. These can help to mitigate the aforementioned challenges and ensure its longevity, sustainability, and impact on the wider quantum community. 

\end{abstract}

\section{Introduction}
The advent of the quantum wave looms over the horizon. Quantum technologies, including quantum cryptography, quantum sensing and quantum computing, harness both superposition and entanglement, which are unique qualities of quantum mechanics, to solve problems beyond the ability of its classical (non-quantum) counterparts~\cite{nielsen2010quantum}. Such improvements may include computational speed-up and the ability to solve currently difficult optimization problems, such as the potential security threat to current encryption standards.~\cite{nielsen2010quantum, wong2023shor}. 

Towards this end, research in quantum technologies have accelerated~\cite{Higgins_Florizoone_Pureswaran_2025, International_2025}. Moreover, because of its potential to improve computational power in various applications spanning a large subset of fields such as machine learning for financial risk modeling, the discovery of new drugs and chemicals, and optimistically the innovation of renewable energies~\cite{cao2018potential, orus2019quantum, giani2021quantum}, there has been an increased interest, and thus funding in quantum technologies, from non-academic sources including various governments. Within Southeast Asia alone, the Philippines' government has pledged to invest ₱860 million (US\$ 14.7 million) by 2030; Thailand's government announced a plan to integrate ฿200 million (US\$6 million) until 2028; and Singapore's National Quantum Strategy, National Research Foundation, and Monetary Authority of Singapore have invested a combined total of S\$800 million (US\$ 624 million) into quantum development~\cite{choong2026quantum,southeast2026manoj, qureca2026, 2026nrf, indonesia2025kompas, philippines2023qpf, philippines2025quantuminsider, thailand2020qpf,thailand2026chiangraitimes,thailand2026thaitimes,malaysia2026nst,malaysia2026oga}.

Concurrently, public engagement in this field has also manifested in quantum grassroots that have sprung up worldwide to support the growth and development of quantum, especially through education~\cite{seskir_democratization_2023, genenz_why_2025}. Grassroots are an important aspect of civil society in terms of being a form of representative democracy for underrepresented communities within the field of quantum technologies, increasing the accessibility of technical information to the general public through open-access and approachable content and involving a more diverse portion of society in this growing field~\cite{seskir_democratization_2023, Garces_Orzechowski_Hernandez_2025, Darienzo_Kelly_Schneble_Wei_2024}. Consequently, grassroots are an important aspect of national quantum movements, yet find little representation in national quantum strategies and policy making~\cite{genenz_why_2025}. 

While there have been comprehensive previous work that explore the structure, goals, challenges, and impact of grassroots which primarily speak English globally~\cite{genenz_why_2025}, there is still a lack of insight into grassroots within the Association of Southeast Asian Nations (ASEAN) region, which are generally younger and still trying to establish themselves. Moreover, few works have explored how government bodies and national strategies can work in tandem with quantum grassroots (and vice-versa). Therefore, this paper aims to provide insight on the ASEAN quantum grassroots initiative from the perspective of various grassroots leaders. Particularly, we identify challenges the fledgling ASEAN grassroots organizations in the region face and the solutions that may benefit and strengthen the longevity of these youth groups.

We have elected to survey grassroots organizations in ASEAN not only due to the lack of coverage of the region, but also its recent and rapid emergence in the quantum technology sphere. Various countries have displayed remarkable efforts in their pursuit of quantum technologies and literary efforts through ventures such as the launch of the BRIN Research Center for Quantum Physics in Indonesia, Philippines' partnership with RIKEN from Japan, and the 2023 launch of the Quantum Technology Research initiative collaboration stemming from Thailand. These developments suggest the region is serious in its investment into the technologies and thus should be supported in the context of its region. Given the fledgling nature of the quantum space in this region, the choice to focus on ASEAN can uniquely provide insights into less established grassroots organisations, which are typically in more tenuous positions to sustain themselves, and how best to support them.

\section{Preliminaries}

\subsection{Quantum Technology}
Quantum technologies have the potential to provide computational speed up and improvements over our current classical computers that can find uses in various fields such as chemistry, computer science, cryptography, sensing and financial modeling~\cite{nielsen2010quantum, orus2019quantum, giani2021quantum}. To do so, quantum technologies leverage on the quantum properties of superposition and entanglement, which in essence enable a type of parallel computing capability that cannot be achieved through non-quantum (classical) computers~\cite{nielsen2010quantum}. Despite being a nascent technology, interest has grown rapidly from both public and private sectors. Not only have investments boomed~\cite{mckinseyinvestments} and the number of patents and papers increased~\cite{quantumindexreport2025}, but governments have also begun to put large amounts of funding into this technology internationally~\cite{qureca2026}.

Its rapid growth can also be seen in the progress in both software and hardware. Various hardware systems for quantum computers are being explored by players such as IBM, Google, and Microsoft~\cite{Jeniecep_2025}. While variants based on superconductors, including that of IBM’s systems having enjoyed the spotlight due to its snowballed progress such as through its exponentially increased computational capacity and error-mitigation techniques over the years~\cite{abughanem2025superconducting}, other quantum hardware systems may instead include photonic, neutral atoms, trapped ions, and quantum dots~\cite{de2021materials, abughanem2024photonic}. While all these systems fundamentally rely on the same two quantum principles of entanglement and superposition, the means through which computational operations are engineered rely on different physical principles depending on their hardware.

Similarly, various quantum software start-up companies such as Horizon~\cite{horizonwebsite}, Quantinuum~\cite{quantinuumwebsite}, and Alice and Bob~\cite{alicebobwebsite}, are engineering different software systems for quantum computation. While these demonstrate the growth of the field, the diversity of platforms also illustrate the varied choice and possible pathways through which quantum technology can continue to develop.

Furthermore, ASEAN's members are beginning to channel more resources into the quantum ecosystem, an example of which follows Table~\ref{tab:funding}. 

\begin{table}[]
    \centering
    \begin{tabular}{|p{3cm}|p{9cm}|}
         \hline
         \textbf{Country} & \textbf{Source of Funds}  \\
         \hline
         Singapore & 2030 Research \& Innovation Plan (US\$28.6 billion) for various uses including acquiring quantum deep tech startups, with Singapore's National Quantum Strategy, National Research Foundation, and Monetary Authority of Singapore investing a combined total of S\$800 million (US\$ 624 million) into quantum development\\
         \hline
         Indonesia & US\$400 million Quantum AI Data Center, combining   ``quantum computing, AI, and big data analytics"\\
         \hline
         Philippines & US\$300,000 quantum computing laboratory for energy-sector implementations in 2025, USD\$14.7 mil Quantum Computing and Artificial Intelligence Research and Development Roadmap in 2023\\
         \hline
         Thailand & Launch of Siam Quantum Square for self-reliant quantum applications, US\$6 million dedicated to the development of quantum technology from 2020-2029\\
         \hline
         Malaysia & Launch of Malaysian Technology Development Corporation (MTDC) Tradeview Quantum Fund with initial capital of US\$4.7 mil and interest to establish National Quantum Council \\
         \hline
    \end{tabular}
    \caption{A sample of some quantum-related funding opportunities present in ASEAN~\cite{southeast2026manoj, qureca2026, 2026nrf, indonesia2025kompas, philippines2023qpf, philippines2025quantuminsider, thailand2020qpf,thailand2026chiangraitimes,thailand2026thaitimes,malaysia2026nst,malaysia2026oga}.}
    \label{tab:funding}
\end{table}

\subsection{Importance and Impact of Grassroots}
\label{sec: importance of grassroots}
In order to continue engaging with quantum technology, a strong foundation is required in various aspects. Complex mathematical and physical concepts underlie its operating principles. Computer science and engineering concepts are required for designing both the software and hardware required to implement and employ quantum technologies ~\cite{nielsen2010quantum}. At a pre-university level, many instructors lack the specific technical knowledge and access to resources to support students in exploring the field of quantum computing, and students at that level may lack an awareness of quantum computing as a potential field. All of these culminate into educational barriers to entry into this field and hinder general public engagement~\cite{Garces_Orzechowski_Hernandez_2025}. Consequently, in addition to national investments in the actual research and development technology, quantum grassroots can and have become an important means towards greater general engagement in the quantum space.

In addition to educational barriers, the work in ~\cite{vishwakarma2024role} found that organizational challenges (such as underrepresentation from minorities) and societal issues were exacerbated in quantum computing due to the perceived high barriers of entry with women and minorities being underrepresented in the quantum field, further belaying the misconception that quantum is only for the ``esoteric and elite." 

\subsubsection{Improving Education \& Perceptions}
The existence of grassroots can help to combat these educational barriers and negative perceptions through multiple avenues, such as the integration of hands-on learning experiences to demystify quantum computing, awareness campaigns, outreach programmes, and encouraging more open access to quantum systems. Such programmes, including for example, IBM's initiative~\cite{vishwakarma2024role}, may help offset the perceived incomprehensibility of the technology and boost public engagement by creating avenues through which the general public can more easily get involved~\cite{seskir_democratization_2023}. In order to broaden the beneficiaries of these outreach programmes, grassroots can also provide virtual events and comprehensive support to not only the general population, but in particular candidates who stem from diverse backgrounds in the quantum space. 

Additionally, outreach programmes by grassroots can leverage on the fact that instructors are often also students or youth themselves, and are thus similar in age to their beneficiaries; `near peer' mentoring in physical sciences have been shown to be beneficial for student outcomes, in terms of performance in school and learning, and retention in science programmes~\cite{zaniewski2016increasing}. This benefit can be attributed to similar aged instructors being more approachable and relatable to students~\cite{Darienzo_Kelly_Schneble_Wei_2024, zaniewski2016increasing}, but also being less likely to suffer from the ``curse of expertise," the phenomenon whereby high technical knowledge in an instructor can lead to less effective pedagogy because they become less attuned to the difficulty and foundations required to learn a concept~\cite{fisher2016curse}. Thus, having grassroots volunteers support ``student-centered, hands-on learning" to address the challenges of abstract quantum concepts can have a positive impact on learning outcomes for beneficiaries of outreach programmes~\cite{Darienzo_Kelly_Schneble_Wei_2024}.

Moreover, educational content can be (and has been) developed with or by youth grassroots, which are often made specifically with younger audiences in mind to explain quantum concepts. These resources are often made open-source (a term generally used to describe software which is freely modified and distributed, though can also be applicable to other mediums of content) and readily available online, which can mitigate regional inequalities by providing quality content for free over the internet~\cite{Garces_Orzechowski_Hernandez_2025}. These resources can also help address the `lagging' nature of education relative to the ``rapid pace of technological development in quantum" to ensure that people are not left behind~\cite{genenz_why_2025}, though it should be noted that obstacles such as language barriers, quality control, and misinformation may lead to the reduction of quality in these programmes~\cite{Garces_Orzechowski_Hernandez_2025}. 

\subsubsection{Representing Underrepresented Communities}
Grassroots can also target organizational and societal issues by uplifting underrepresented communities. Grassroots allow for the perspectives and opinions of its specific target demographic to be represented in decision-making, particularly in the progression of quantum technologies. In addition to minority groups such as women in quantum, grassroots can also help to represent consumers, who likely will not be able to voice an opinion in the development of quantum technologies until they are affected by its results. Therefore, grassroots can, with their informed perspectives, represent the voices of frequently underrepresented demographics ~\cite{seskir_democratization_2023}.

\subsubsection{Shaping Technological Direction}
The integration of quantum grassroots can have broader impacts on the quantum community and the direction of research undertaken within the community. In ~\cite{gercek2025navigating}, they highlight the term path-dependency, explained as ``decisions of the past put[ting] a constraint on the range of decisions available and their cost today, and the decisions today [putting] a constraint on the range of available decisions and their cost in the future." Essentially, the possible trajectories of future research and development will naturally be constrained by the choices of which technologies are favoured and adopted in the past and present.

Contextualizing this to quantum technologies, superconducting computers are at the forefront of quantum technology development (with  Google's superconducting-based processor, Willow, being the first to demonstrate hardware error correction and independently verifiable quantum advantage~\cite{bbcGoogleWillow}), despite its weaker performance in noise susceptibility compared to its other hardware counterparts~\cite{humble2019quantum}. This focus is primarily due to its cost-efficiency and the high design control it affords over other technologies. The current booming focus on superconducting technologies will naturally lead to more funding and research to be undertaken within those fields, with ion-based and photonic-based research (among others) falling to the side. Naturally, this focus will strengthen the development of superconducting qubits, and will likely continue to cement its place at the forefront of the technological race. Consequently, it will only become more expensive for other possible hardware systems to develop and catch up to the level of superconducting circuits. This essentially risks ``locking" research direction into a specific pathways (i.e., superconducting-based technologies), at the cost of other potential developments.

This may become a problem given that quantum technologies are still young, and it is unclear which hardware type will ultimately supersede the others in terms of cost, scaling and computational prowess. There is also the possibility that the `optimal' choice of technology is highly reliant on different budgets, use cases, and the target demographic, meaning that multiple paths of development that should be explored to best suit different constraints. It is therefore essential to avoid this trap of path-dependency too early and allow all possible technologies to continue to develop. Thus, grassroots can help to explore the beginnings of these different pathways to ensure that various potential hardware are introduced fairly and explored, though it should be recognized that the extent of its effectiveness is also dependent on the amount of manpower available and the expertise and/or skill-sets of those involved. Nevertheless, the collaborative nature of grassroots can act as a means to support a more diverse exploration of these different pathways.

\subsubsection{Fostering Network, Community and a Quantum Workforce}
Grassroots may also play a role in creating opportunities for students to learn and acquire careers in the field on a global scale. Through outreach and awareness programmes in collaboration with quantum companies, grassroots may act as intermediaries that connect these companies with local talents. Grassroots also play a part in the development of a community and a quantum ecosystem where the presence of quantum technologies is not yet cemented ~\cite{genenz_why_2025}.

\section{Methodology}

\subsection{Surveying and interviewing different grassroots organizations}

In order to understand the perspectives of grassroots organizations in ASEAN and the challenges they face, we conducted two phases of investigation. In the first phase, we began with a a survey with questions focused on three major aspects of their organisation: (1) organizational goals and the steps taken to align with these goals; (2) key actions they wish to implement but have not yet done so; and (3) current issues and obstacles that are preventing these organizations from implementing the aforementioned key actions. We also queried them on the type of resources and solutions they felt would assist them in navigating or mitigating these limiting factors.

We received a total of ten responses representing seven different grassroots organizations of the following five ASEAN countries: Indonesia, Malaysia, Philippines, Singapore, and Thailand. A list of all the questions we asked can be found in the Appendix. The names and responses of the organizational representatives have been anonymized. 

Following this preliminary data collection, we conducted more qualitative email interviews with multiple survey respondees in order to deepen our perspective on their responses. These interview questions were tailored to each grassroot's representative(s) based on their original survey responses. A list of the follow-up questions and the countries can be found in the Appendix.

\section{Results and Analysis}

\subsection{Organizational Goals}
Broadly, five out of seven of the organizations mention either democratizing quantum literacy, increasing access to knowledge, education, or a combination of those three factors as part of their organisational goals. Within that five, three additionally reference grassroots' symbiotic relationship with the workforce in some capacity: one making quantum opportunities more accessible to youths, while two make explicit references to building a talent pipeline and a quantum-literate workforce. Accessible education is an important aspect in building a future workforce~\cite{vishwakarma2024role}, particularly for students, as it narrows the divide caused by the relatively high barriers to entry of specializing in quantum technologies ~\cite{Garces_Orzechowski_Hernandez_2025, gercek2025navigating}.

Within the remaining two organizations who do not have education or quantum literacy mentioned explicitly in their goals, one mentions ``building [a] quantum research community" and ``establish[ing] a sustainable and impactful quantum technology ecosystem" in their home country.

For the last remaining organization, its goal is primarily to serve as a connective tissue between academic research and real-world implementation, citing they aim to ``apply deep technical knowledge to create practical solutions in many different fields."

In fact, one other response posits their grassroots to take on a similarly more technical research role as compared to other surveyed grassroots, stating their overarching goal is to transition their country from a ``technology consumer to a sovereign producer." To this end, for the latter organisation, one of their goals is to have ``applied `mission-driven' research rather that pursuing theoretical research in isolation" and to ``provide the physical tools necessary for a self-sustaining ecosystem" to ensure ``infrastructure \& technical sovereignty." Additionally, they structure themselves as a ``solutions-driven research organization" which aims to locate the needs of various industries, with examples in applied engineering and industrial management, to source a ``deep understanding of both the traditional industrial landscape and fundamental technology." Notably, this organization is neither that of a conventional grassroots, nor is it entirely dedicated to quantum --- thereby highlighting the diversity of grassroots organizations within the same space. As such, we include their responses to provide more varied perspectives on how support can contextually be disseminated to each grassroot community.

Finally, one of the five organizations that highlights quantum literacy also underscores their desire to ``support grassroots initiatives through cross-pollination and collaboration" as well as to ``amplify the collective impact of grassroots efforts by increasing their visibility and enabling stronger support in the future," thereby signalling its interest in occupying a broader and `higher' nurturing role within the quantum grassroots communities.

Therefore, across the board, these quantum grassroots demonstrate in their goals an awareness of its role within a wider network of governing bodies, professional industries, academics, and even other quantum grassroots internationally, despite the difference in the extent of their desired engagement with these other players in the field. However, it is also evident from these responses that not all grassroots work towards the same goal: while most aim to cover either an aspect of community-building or education, some are also interested in covering a more technical and research-forward role.

We note also that through the goals of these grassroots, they cover most of the cited impacts of grassroots organizations in Section ~\ref{sec: importance of grassroots}, primarily improving education and fostering a community.  However, only two of seven of the grassroots have an explicit interest in shaping the technological direction of quantum research in their country.

\subsection{Targeted Demographic of the Organizations}
Despite a varied current targeted demographic across all grassroots organizations, across {\it all seven} of these organizations, the common target demographic mentioned are students. Three out of seven organizations specified university students in particular, while one highlighted that their only targeted demographic thus far has been high school students. Despite that, the latter response also indicated an interest to expand their target demographic to university students. This strongly positions quantum grassroots in ASEAN as a channel to access, connect and nurture young talents who could enter the workforce in the short-term.

However, the grassroots surveyed also have a keen interest in connecting with three other key players in the ecosystem: (1) scientists and academics, (2) industry leaders (such as companies, venture capitalists and startups), and (3) national government officials and policymakers. In ~\cite{genenz_why_2025}, they found that quantum grassroots in ``countries with lower investments and awareness of [quantum technologies]" tended to shift their target demographic to these other groups. While there is information on the approximate funding sizes for quantum in ASEAN (see Table~\ref{tab:funding}) this data may be insufficient to be fully representative of all the funding in the region to allow us to draw any conclusions along these lines.

For scientists and academics, four out of seven organizations have them as their target demographic. The corresponding goals of these four organisations were related to community-building and empowering applied research, both of which might explain this chosen target demographic.

For industry leaders and companies, two out of seven already have them as part of their target demographic, while three out of seven explicitly list them as their intended expanded demographic. Again, both of the two whose target demographic includes this group of people have organisational goals related to community-building. Of the three who intend to expand, their goals are more varied. Their goals, separately, are: empowering applied research; building a research community; and making quantum more accessible for youth, including in terms of network and opportunities.

Finally, for national government officials and policymakers, two out of seven list them as their current target demographic. Moreover, three out of the seven grassroots mention strengthening ties with this demographic as future steps for their organization: one responder ``hope[s] to build research network" of ``scientists, students, industries, governments." Another responder mentions they want to continue to ``grow [their] engagement with research, policy, industry, and quantum-safe cybersecurity." The last responder added that part of their grassroots' next steps is to ``[align] quantum development with national needs like energy sovereignty." These responses clearly represent a strong interest within some of the grassroots organisations to align with policy makers and government bodies. Therefore, while  none of the grassroots explicitly noted their goal to represent underrepresented communities in decision-making, their interest in engaging with policy makers and governing bodies seem to imply an interest to participate in these conversations, thereby motivating a need to look towards how the two sides to work in tandem towards a quantum-ready future.

\subsection{Next Steps for the Organizations}

Four out of seven organizations indicated they would like to aim to, in some capacity, strengthen their community ties in terms of more engagements with other parties, bettering relations with other grassroots organizations, and expand research networks. One organization aims to reach more communities across various regions of their country to ``help build a stronger national quantum ecosystem," while another aims to ``[m]aintain the strong ties [they] have made from the different quantum ecosystem players." One responder also highlighted their interest to ``build research network and stronger communities of quantum scientists in ASEAN region," aligning with their interest to expand their target demographic to other ASEAN countries. All of these depict quantum grassroots' aim to ensure that the community benefits from their outreach and their understanding that fostering collaborations, both locally across different players in the quantum ecosystem, or internationally, amongst different grassroots organisations, is key to this end.

Some organizations would like to achieve better infrastructure and internal processes. This includes the improvement of infrastructure, the recruitment of more manpower, and bettering internal processes such as through ``documentation, program management, and continuity, while also securing more sustainable funding to help us initiate and scale projects," as one organization has phrased it. The organization whose goals also include supporting and nurturing other quantum grassroots communities aims to ``properly establish itself into a non-profit organization." It is clear that these organizations wish to establish themselves as key players in the ecosystem and ensure their longevity by working towards a more sustainable workflow to not only be able to maintain, but also scale up their operations to aid their missions.

In general, most grassroots organizations also aim to produce more output and increase their impact, though the extent to which they wish to do so varies depending on their targeted demographic and goals. Towards the educational and outreach-centric goals, grassroots have hopes to deliver events, expand their educational efforts, and translate technology into educational use-cases. Three grassroots also explicitly mention having international influence and collaborations as part of their ideal next steps.

More ambitious grassroots, particularly those with interests to shape the technological development of the research in their country, have also set their sights on influencing the local economy. One organization has expressed its next step to ``[position] the organization as a strategic receiver that translates advanced technology and IoT solutions into practical industrial use cases," while another has mentioned their desire to ``mov[e] from theory and basic research into a tangible, national-scale infrastructure." ``Creat[ing] a tangible result for the national economy using quantum technologies" is in the cards for another grassroots organization, and to this end, they wish to ``grow [their] engagement with research, policy, industry, and quantum-safe cybersecurity." Evidently, quantum grassroots organizations wish to have a ready impact on the exploration into new quantum technologies and the potential first-steps for the younger generations to foray into this novel technology. 

\subsection{Obstacles to Achieving Results}
Despite the ambitions of these grassroots organizations, they have to contend with limitations and constraints that can extenuate  their impact.

The biggest fragility point for quantum grassroots organizations is manpower, particularly those who have technical or specialized quantum skills. With an organization having only ``a few committed volunteers helping drive the work" and having ``so much depend[ing] on a few people who are also balancing professional and personal responsibilities," it typically leads to lacking manpower which makes it difficult for organizations to scale up in their efforts. For example, one responder mentions that the phenomena has led to "burnout, volunteer churn, and [an increased amount of] leadership transition."

Another organization highlights what they define as ``aspirational churn" being having mismatched expectations based on the amount of effort they will have to take in more foundational subjects such as mathematics and physics before they are allowed to ``even touch a quantum processor." As talent is scarce but the demand high, and people who take interest comprised of those who have the existing, difficult to build technical knowledge that takes much time and effort, the pool of capable, available talent is small and that could lead to ``burnout [and] volunteer churn."

Three out of seven organizations acknowledge one of their biggest fragility point is funding, and five out of seven organizations either hint at or directly acknowledge a limitation to achieving their goals is the lack of funding. One respondent additionally commented that both ``funding and burnout" is ``expected for voluntary works." Having a source of financial income allows for participants to, as one organization notes, ``move from ideas and plans to actual implementation, while also reducing the strain on our small volunteer base." An organization stressed they intentionally planned events that ``didn't require much funding," which could have led to limited outreach and impact on the community. The statement implicates that the lack of financial support required to ensure plans are enacted could lead to various organizations being unable to have a positive impact on the community, as many desire to achieve.

Interestingly, a few organizations mention difficulty in ``navigating [the] regional regulations" and ``obtaining top-down support," with one citing ``the complexity of legal documentation and the consent process" as a primary challenge, implicating the various paperwork and bureaucratic hurdles they have to navigate in order to actuate their plans. In particular, this organization stresses how it is difficult to have practical implementations of their developments enacted cross-borders due to the ``sensitive requirements" of different jurisdictions, thus requiring ``careful... evaluat[ion] before broader deployment." This thereby highlights how current legislative support may still be opaque. Barriers such as the lack of understanding as to which documents are required, the inadequate support from institutions who can make legislative changes, and the sheer daunt of the many required steps can deter grassroots from wanting to attempt any practical quantum implementations due to concerns over the waiting time or legal repercussions should they incidentally and unintentionally miss important documentation and permissive steps.

Along similar lines, one grassroots organization also highlights that quantum, being an emerging field, has lead to ``limited local expertise, infrastructure, and institutional readiness." Therefore, it is clear that there are also institutional or regulational barriers that quantum grassroots in the organisation still has to contend with.

Some grassroots also mention that, due to their infancy, they are currently in the ``exploration phase" and may not be aware of specific impending obstacles. We believe this statement may apply more broadly to the other organizations as well. For example, none of the organizations brought up potential challenges with leadership transitions as members began to leave the organization, or contending with hype surrounding the technology mentioned in other studies~\cite{genenz_why_2025}. Therefore, this section may not be an all-encompassing list of struggles that quantum grassroots will encounter, but broad categories of issues they have highlighted in their journey thus far.

\subsection{Towards Supporting Grassroots}
It is clear that grassroots organizations can have a significant impact on the quantum ecosystem, and that current grassroots organizations within ASEAN themselves have ambitions to occupy a substantive role in the community. It is also clear that the progress and sustainability of these grassroots organizations are hindered by obstacles explored in the previous section. It is thus important for other players in the ecosystem to also look towards better supporting grassroots organizations in a manner that can mutually benefit all stakeholders involved. Here we explore some potential avenues of support through which grassroots could stand to benefit from in order to achieve their potential to strengthen ties with their communities, overcome educational barriers, demystify quantum technologies, and lead to the creation of more opportunities for career paths in this field. 

\subsubsection{Access to mentorship and expertise}
Grassroots organizations may benefit from the expertise and know-how with access to the advice of local talent, such as academics and industry leaders, with specialized knowledge in various quantum technologies. This can help organizations, which have the goal of generating educational content, with the process of content creation and verification. Whether this is through the active involvement of technical experts in the creation of content, or through a consulting role, access to technical expertise can help mitigate the issue of misinformation~\cite{Garces_Orzechowski_Hernandez_2025} and heighten the quality of their content. In return, for industry players, this can also have the benefit of a more widespread adoption of their technologies and develop future talent that is well-versed in important scientific and technical concepts.

Additionally, with most educational quantum content being taught in English~\cite{Garces_Orzechowski_Hernandez_2025, seskir_democratization_2023}, countries whose residents primarily communicate with other first languages may not assimilate the nuance of the educational content as well as they could with languages they are more familiar with. As such, access to the mentorship of trained professionals fluent in the nuances of some quantum technologies, English, and their mother tongue can greatly benefit in ensuring the quality and accuracy of translated content. All of these will work towards lowering the educational barrier and developing a more quantum-ready society and talent pool.

However, mentorship need not purely be localized in terms of technical quantum expertise. With regards to educational content, grassroots organizations can also benefit from pedagogical instruction, helping them to become better educators in not only the content they create, but also in outreach and awareness programmes.

Moving towards more organizational concerns, mentorship may also greatly benefit grassroots organizations in streamlining their workflow and better managing their organization. This knowledge can greatly benefit grassroots who struggle with managing the organization of manpower and fast turnover rate and help to mitigating issues when leadership transitions arise. Teaching members of the organizations these skills will not only benefit the organizations, but may carry across to careers of members who have left when they settle into other occupational roles in the future. 

\subsubsection{Openness to collaboration}
With all surveyed grassroots organizations indicating that they want to expand their outreach across communities, connecting organizations with decision-makers and key players in both the industry and governance can assist in the ``acqui[sition of] more support for international collaborative projects" and ``increas[ing] engagements with various adjacent and technical places."

Connections may allow for the conception of innovative projects, such as cross-industry collaborations as one initiative seeks to achieve, instead of having to worry about institutional research speeds and balancing it with bureaucratic processes (for instance, the slow acquisition of hardware). Additionally, these collaborations can also leverage already-existing connections between grassroots organizations which can help heighten the reach of impact of these collaborative projects.

Both connections and collaborations can be helpful in engagement and outreach programmes for the wider community by fostering a connection between industry leaders, academics and the general public. This can not only promote potential multi-stakeholder projects, but also portray quantum as a more relevant and approachable topic to audiences with little exposure.

Collaboration between decision-makers and governance may also help to reduce the friction between regulators and grassroots, who may find difficulty navigating institutional limitations. By allowing for a better channel of communication, resolutions between conflicting interests may be resolved quicker and more consistently. For example, through iterative and constructive communication in the long term amongst relevant stakeholders, clearer documented instructions for bureaucratic processes can be developed. As mentioned, current limitations surrounding institutional support on legislation may hinder grassroots organizations from going beyond their expected `educational' scope\footnote{This is not to say grassroots organizations are expected to mainly teach, but as is our observation a majority of the organizations had education as one of their higher priorities. As such, this move allows them to broaden their horizons.}; changes towards this can aid grassroots in producing more practical quantum solutions beyond educational content. The provision of more transparent documented instructions can encourage grassroots organizations to take on more challenges and diversify their services, while aligning with and achieving their other goals of spreading awareness of quantum technologies and educating the youths. Achieving practical experience in an already arguably gated field is daunting, and in a topic which already has such a high barrier to entry, one more step --- even administratively --- made easier is progress which will help grassroots' enter into this foray more confidently.

Finally, bilateral ties between industry and grassroots organizations may also open up avenues for both parties to benefit by providing a channel through which young talent may find job or internship opportunities.

\subsubsection{Access to financial assistance}
Four out of seven grassroots organizations have, either as a fragility point or a limitation to achieving their goals, listed the lack of funding as an issue. They typically rely on volunteers and passionate students who work for little to no monetary compensation. However, the provision of financial support will allow these grassroots organizations to conduct improvements to both their operative processes and the scope of their reach by hiring full-time administrative support and allow these organizations to afford resources, such as proper compensation for creators, content editors, and researchers who volunteer and/or offer their expertise, to reach communities. Even pooled financial support not dedicated only to grassroots which provide them some allocated resources can benefit these organizations. With this in effect, grassroots organizations can reach more systemic, consistent, and sustainable internal operations. The focus can then be shifted towards the implementation of plans as opposed to primarily worrying about manpower and resource constraints. 

\section{Limitations}
While this study has provided more insight into quantum grassroots in ASEAN and how to better support them, it has its limitations. Firstly, not all ASEAN countries were represented in this study, with the omission of notable players such as Vietnam, Brunei, and Cambodia. Moreover, responses were received and aggregated in English, which may have resulted in lost nuances of those whose first languages are not English. While we have attempted to mitigate these through a digital form and follow-up email interviews to clarify certain points, both of which provided the opportunity for respondees to fine-tune their responses, it is possible that some fine-grained distinctions may still not have been fully encapsulated.

Responses were taken from organizational heads or representatives who have provided their opinions on their quantum grassroots' growth, struggles, opportunities and limitations. However, as a maximum of two respondents per organization were taken, it may not be entirely indicative of an entire organization's position. Opinions may vary based on the roles respondents are in, responsibilities, workloads, and other relevant factors. 

\section{Conclusion}

In this paper, we outlined the horizon of quantum technologies and its potential applications as well as grassroots organizations, their impact, benefits, and the distinct lack of coverage within the ASEAN region. We identified their varied goals that cover building a community (both locally and internationally), improving quantum literacy, and shaping the technological direction of quantum technology development within their country. However, we also identified major obstacles including the lack of manpower, the lack of funding, and institutional and regional regulations. To help grassroots organizations realize their potential to be key long-term players in the quantum space, we provide suggestions as to how these organizations can be better supported. We suggest that they will benefit from the expertise and know-how from local talent with specialized knowledge, more connections to key stakeholders in the quantum technology space, and the provision of funding. With the support they require, the hope is that these grassroots organizations will be able to bridge the gap between local communities, assist in the exposure and basic education of quantum technologies, and be part of shaping a quantum-ready workforce.

\section{Acknowledgements}

We would like to thank Dr. Zeki C. Seskir for the helpful discussions surrounding how to approach to this paper. We would also like to thank Kai Tze Tam for his administrative support and his assistance in reaching out to the various grassroots organization heads for the purposes of data collection. 

\printbibliography

\newpage
\section{Appendix}

\subsection{Grassroots survey questions}

\begin{table*}[hbt!]
    \centering
    \begin{tblr}{|p{0.7\linewidth}|p{0.3\linewidth}|}
         \toprule
         \textbf{Question} & \textbf{Response type} \\
         \midrule
         What is the name of the organization you represent? & Short answer text \\
         \midrule
         What is your name? & Short answer text \\
         \midrule
         What are the goals of your current organization? & Short answer text \\
         \midrule
         To what extent do you think your current organization is achieving those goals now? & Short answer text \\
         \midrule
         What are actions/activities that your grassroots organisation is enacting in order to align with your goals? & Long answer text \\
         \midrule
         Who is the targeted demographic of the outreach that you do? & Short answer text \\
         \midrule
         Prertaining to the question above, are there interests to expand targeted demographic? & Short answer text \\
         \midrule
         What are the next steps for your organisation? & Long answer text \\
         \midrule
         Are there things that you would ideally wish to implement or improve on in the workflow and scope of your organization? & Long answer text \\
         \midrule
         Are there limitations that are currently preventing your organisation from implementing these? If so, what are they? & Long answer text \\
         \midrule
         What's the biggest fragility point for grassroots in your context (e.g., funding, volunteer churn, leadership transition, burnout)? & Long answer text \\
         \midrule
         Where does your organization struggle most (e.g., speed, outreach reach, trust, tailoring to niche groups)? & Long answer text \\
         \midrule
    \end{tblr}
    \caption{Surveyed questions and their corresponding answer types}
\end{table*}

\end{document}